\documentclass[twocolumn,showpacs,oneside,amsmath,amssymb,prl]{revtex4}
\usepackage{amsmath}
\usepackage{graphicx}
\usepackage{epstopdf}
\usepackage{dcolumn}
\usepackage{bm}

\begin{document}

\title{Mott Physics and Topological Phase Transition in Correlated Dirac Fermions}

\author{Shun-Li Yu$^1$}
\author{X. C. Xie$^{2,3,4}$}
\author{Jian-Xin Li$^1$}
\affiliation{$^1$National Laboratory of Solid State
Microstructures and Department of Physics, Nanjing University,
Nanjing 210093, China}
\affiliation{$^2$International Center for Quantum Materials, Peking University, Beijing 100871, China}
\affiliation{$^3$Institute of Physics, Chinese Academy of Sciences, Beijing 100190, China}
\affiliation{$^4$Department of Physics, Oklahoma State University, Stillwater, Oklahoma 74078}

\date{\today}

\begin{abstract}
We investigate the interplay between the strong correlation and the spin-orbital coupling in the Kane-Mele-Hubbard model and obtain the qualitative phase diagram via the variational cluster approach. We identify, through an increase of the Hubbard $U$, the transition from the topological band insulator to either the spin liquid phase or the easy-plane antiferromagnetic insulating phase, depending on the strength of the spin-orbit coupling. A nontrivial evolution of the bulk bands in the topological quantum phase transition is also demonstrated.

\end{abstract}

\pacs{03.65.Vf, 71.27.+a, 71.10.Pm, 71.30.+h}

\maketitle

In recent years, a new field has emerged in condensed
matter physics, based on the realization that a
spin-orbit interaction can lead to topologically insulating
electronic phases~\cite{Hasan,Qi}.
A topological band insulator (TBI) has a nontrivial
band structure resulting from the strong spin-orbit coupling.
Theoretical and experimental studies have found such materials
in both two (2D)~\cite{Kane,Bernevig,Konig} and three (3D)~\cite{Hsieh1,Hsieh2,Xia,Zhang,Chen} dimensions.
A common property of TBI is that it has a charge excitation gap in the bulk,
but with gapless helical edge (or surface) states protected by the time reversal
symmetry lying inside the bulk insulating gap. As a new quantum state, which is the $Z_{2}$-graded topological distinction from other
conventional insulators, it has attracted great attention.
Though great progress has been achieved, the current researches mostly focus on the weakly interacting
systems. It has been proposed that the topological insulator may also appear in the systems with substantial electron
correlations, such as $4d$ and $5d$ transition metal oxides~\cite{Pesin,Shitade}. And the electron interaction effect plays a crucial role in determining the ground state of topological insulators in the 2D limit~\cite{Wang}. Therefore, the effects of electron correlations on the topological insulators present a new challenge.

The correlation effects in topological insulators can be studied either by interaction-driven
topological insulators~\cite{Raghu,Sun,Dzero} or by introducing
interactions to a system with a strong spin-orbit coupling~\cite{Pesin,Rachel,Varney}. In this Letter, we investigate
the model proposed by Kane and Mele~\cite{Kane} on the honeycomb lattice for describing a 2D topological
insulator, and introduce the Hubbard interaction to this model to analyze the Mott physics.
Recently, the Hubbard model on the honeycomb lattice have been studied by Meng \emph{et} \emph{al}~\cite{Meng}
using the quantum Monte Carlo (QMC)
method, in which a spin liquid (SL) phase is found to exist between the semi-metallic (SM) phase and
the antiferromagnetically (AF) ordered Mott insulator (MI) phase for a range of the on-site interaction $U$.
The mean field analysis and QMC simulations for the Kane-Mele-Hubbard (KMH) model reveal that the TBI phase is unstable against the magnetic ordering phase~\cite{Rachel,Hohenadler,Zheng}.
But the whole phase diagram of the KMH model, especially the transition between the TBI and the MI, and the nature of the single-particle excitations in the bulk and on the edges  are still open theoretical questions. As the existence of gapless edge states
is the direct manifestation of the topological
nature, the study of the single-particle excitation spectra is the natural way to investigate the phase transition between TBI and MI.
Here, we use the (zero temperature) variational cluster approach (VCA)~\cite{Potthoff}, which goes beyond the mean field theory and takes into
account exactly the effects of short-range correlations by an exact diagonlization of the separative clusters. We find a topological quantum phase
transition from TBI to MI with increasing
$U$ and this process shows a nontrivial evolution. Starting from TBI, the spin-orbit coupling gap $\Delta_{SO}$ closes first and then the Mott gap opens up but without the gapless edge states for increasing
$U$, which is closely related to the topological properties of the system. The closing process of $\Delta_{SO}$ driven by the correlations is accompanying with a splitting of both the conduction and valence bands.
In the strong spin-orbit coupling regime, the state transiting from TBI is the easy-plane AF Mott insulator. In the weak coupling regime, a spin liquid phase emerges between the TBI and the AF Mott insulators.
In addition, we also find a decrease in the velocity of the helical edge states
due to the correlations in the TBI phase.

\begin{figure}
  \centering
  \includegraphics[scale=0.3]{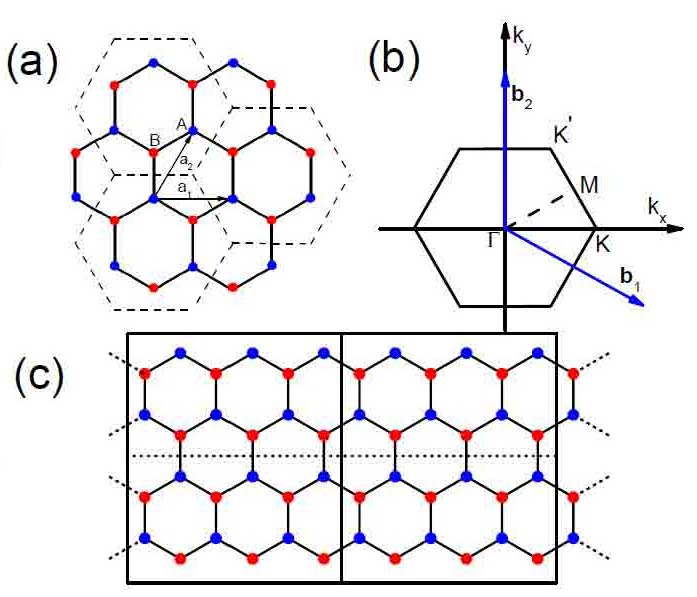}
  \caption{(color online) (a) The 6-site cluster tiling (dashed lines) on honeycomb lattice used in our calculations of the bulk properties with VCA.
  A and B denote the two inequivalent sites, $\mathbf{a}_{1}$ and $\mathbf{a}_{2}$ are the lattice unit vectors.
  (b) The first Brillouin zone. $\mathbf{b}_{1}$ and $\mathbf{b}_{2}$ are the reciprocal-lattice vectors.
  (c) Schematic view of tiling the ribbon for calculating the edge states. Here, two superlattices (rectangle with solid lines ) are shown. Each superlattice
  makes up of two clusters containing 12 sites as separated by the dashed line).}
  \label{fig1}
\end{figure}
The Kane-Mele-Hubbard model is defined as $
H=H_{0}+H_{U}$, where $H_{0}$ is the model proposed by Kane and Mele
on the honeycomb lattice as shown in Fig. \ref{fig1}(a)~\cite{Kane},
\begin{eqnarray}
H_{0}=t\sum_{\langle ij\rangle\sigma}c^{\dag}_{i\sigma}c_{j\sigma}+i\lambda\sum_{\langle\langle ij\rangle\rangle\sigma\sigma^{\prime}}
\nu_{ij}c^{\dag}_{i\sigma}\tau^{z}_{\sigma\sigma^{\prime}}c_{j\sigma^{\prime}},
\label{eq1}
\end{eqnarray}
and $H_{U}$ the Hubbard interaction,
\begin{eqnarray}
H_{U}=U\sum_{i}n_{i\uparrow}n_{i\downarrow}.
\end{eqnarray}
Here, $\langle i,j\rangle$ and $\langle\langle ij\rangle\rangle$ denote the nearest neighbor (NN) and the next nearest neighbor (NNN), respectively. $\lambda$ is the spin-orbit coupling constant and $\tau$ the Pauli matrices.
$\nu_{ij}=+1(-1)$ if the electron makes a left(right) turn to get to the NNN site. Others are in standard notation.

The VCA is a cluster method of the self-energy functional approach (SFA)~\cite{Potthoff}, which approximates the self-energy of the original system by the self-energy $\mathbf{\Sigma}^{\prime}$ of an exactly
solvable reference system with the same interaction term. It has been successfully
applied to, for instance, the problem of competing phases in high-$T_{c}$ superconductors~\cite{Senechal,Aichhorn}. Despite the
considerable finite-size errors, the VCA can predict the
qualitatively correct trend for the phase diagram~\cite{Balzer}. In VCA, the lattice is tiled into identical clusters (as illustrated in Fig. \ref{fig1}, each cluster contains a hexagon in our calculations) and the reference system is made up of the decoupled clusters.
The single-particle parameters (denoted by $\mathbf{t}^{\prime}$) of the reference system are optimized according to the variational principle.
And one can add any Weiss field to study the symmetry broken phases.
For any $\mathbf{\Sigma}^{\prime}$ parameterized as $\mathbf{\Sigma}^{\prime}(\mathbf{t}^{\prime})$, we have the grand potential:
\begin{eqnarray}
\Omega[\mathbf{\Sigma}^{\prime}(\mathbf{t}^{\prime})]&=&\Omega^{\prime}(\mathbf{t}^{\prime})+\mathrm{Tr}\ln[-(\mathbf{G}_{0}^{-1}-
\mathbf{\Sigma}^{\prime}(\mathbf{t}^{\prime}))^{-1}] \nonumber \\
&-&\mathrm{Tr}\ln[-\mathbf{G}^{\prime}(\mathbf{t}^{\prime})],
\end{eqnarray}
where $\Omega^{\prime}(\mathbf{t}^{\prime})$ and $\mathbf{G}^{\prime}(\mathbf{t}^{\prime})$ are the grand potential and Green's function of the reference system, $\mathbf{G}_{0}$ is the free Green's function
without interactions. The physical self-energy $\mathbf{\Sigma}$ is given by the stationary point $\partial\Omega[\mathbf{\Sigma}^{\prime}
(\mathbf{t}^{\prime})]/\partial \mathbf{t}^{\prime}=0$. For any $\mathbf{t}^{\prime}$, $\mathbf{\Sigma}^{\prime}(\mathbf{t}^{\prime})$ is related to the lattice Green's function
$\mathbf{G}(t^{\prime})$ by the Dyson equation $\mathbf{G}^{-1}(\mathbf{t}^{\prime})=\mathbf{G}_{0}^{-1}-\mathbf{\Sigma}^{\prime}(\mathbf{t}^{\prime})$. $\mathbf{G}(\mathbf{t}^{\prime})$ can be determined
via the cluster perturbation theory~\cite{Senechal1}, in which $\mathbf{G}^{\prime}(\mathbf{t}^{\prime})$ is calculated by the exact diagonalization method and the intercluster hopping $\mathbf{V}$ is
treated perturbatively. In momentum space, $\mathbf{G}(\mathbf{t}^{\prime})$ can be expressed in terms of $\mathbf{G}^{\prime}(\mathbf{t}^{\prime})$ and $\mathbf{V}$ as $\mathbf{G}(\mathbf{k},\omega)=\mathbf{G}'(\mathbf{k},\omega)
[1-\mathbf{V}(\mathbf{k})\mathbf{G}'(\mathbf{k},\omega)]^{-1}$.

To calculate the edge states, we consider a strip geometry and construct a supercluster
which is made of several clusters. In Fig. \ref{fig1}(c), for example, we arrange two clusters (12 sites in one cluster) in $y$ direction to form a supercluster.
The Green funciton of the supercluster is given as $(G^{sc})^{-1}=(G^{\prime})^{-1}-W$ and
$W$ is the intercluster hopping matrix in the supercluster.

To test the existence of the possible AF order, we will include the following
Weiss field,
\begin{eqnarray}
H_{AF}^{\alpha}=h^{\alpha}_{AF}\sum_{i}(-1)^{\eta_{i}}c^{\dagger}_{i\sigma}\tau^{\alpha}_{\sigma\sigma^\prime}c^{\dagger}_{i\sigma^\prime},
\end{eqnarray}
where $\eta_{i}=0$ or 1, when $i\in A$ or $B$.
In the absence of the spin-orbit interaction, the spin sector has a $SU(2)$ symmetry. So, we have $h^{z}_{AF}=h^{x,y}_{AF}$. However, this relation is broken down when the spin-orbit interaction is turned on. In this case, we will calculate the grand potential $\Omega(h_{AF})$ as a function of $h^{z}_{AF}$ and $h^{x}_{AF}$, respectively.

\begin{figure}
  \centering
  \includegraphics[scale=1.0]{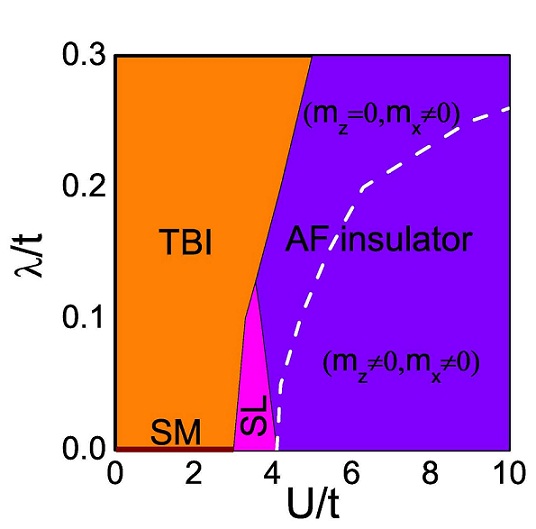}
  \caption{(color online) Qualitative phase diagram of KMH model.
  SM, TBI, SL and AF insulator denote the semi-metal, topological band insulator, spin liquid and
  antiferromagnetic insulator, respectively. Above the white dashed line in the AF insulator phase,
  the $z$-term of the AF order disappears. }
  \label{fig2}
\end{figure}

\begin{figure}
  \centering
  \includegraphics[scale=0.4]{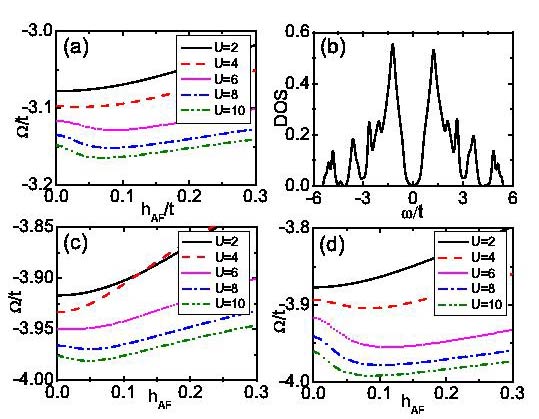}
  \caption{(color online) (a) $\Omega$ as a function of $h_{AF}$ for various values of $U$ at $\lambda=0$.
  (b) The density of states for $U=4$ and $\lambda=0$. (c) and (d): $\Omega$
  vs $h_{AF}$ at $\lambda=0.2t$ along the $z$ and $x$-directions, respectively.}
  \label{fig3}
\end{figure}

Our main results on the interplay between the Hubbard interaction and the spin-orbit coupling are summarized in the $U-\lambda$ phase diagram [Fig. \ref{fig2}]. Let us first discuss the $\lambda=0$ line. In VCA, the existence of the AF order can be determined by the $h^{\alpha}_{AF}$ dependence of the grand potential $\Omega(h_{AF})$. Fig. \ref{fig3}(a) presents the results for different Hubbard interactions $U$. For weak $U$, such as $U=2t$ and $4t$, $\Omega(h_{AF})$ shows a monotonic increase with $h^{z}_{AF}$ ($h^{z}_{AF}=h^{x}_{AF}$ in this case), indicating that no AF order forms in the system. However, for a large $U$ such as $U\geq6t$, a minimum appears at finite $h^{z}_{AF}$ and this minimum moves to lower $h^{z}_{AF}$ values with increase of $U$. Therefore, we can infer that an AF order exists for a large $U$ as expected. Interestingly, when plotting the density of states (DOS) for $U=4t$ as shown
in Fig. \ref{fig3}(b), we find that an obvious Mott gap has opened up around the Fermi energy. This paramagnetic insulating phase is identified as the SL phase as also been found
recently by Meng \emph{et} \emph{al} using the QMC simulation~\cite{Meng}. Therefore, the system will undergo phase transitions from the semi-metal(SM) to SL and then to AF Mott insulator with the increase of $U$. Thus, we can reproduce the QMC simulation results calculated for $\lambda=0$~\cite{Meng}.

When turning on the spin-orbit coupling, we find that the SL phase maintains for a range of spin-orbit coupling up to $\lambda=0.125t$. On the other hand, the AF order is not isotropic. As seen from Figs. \ref{fig3}(c) and (d), no minimum is found at $U=4t$ for $\lambda=0.2t$ in the $h^{z}_{AF}$ dependence, but it can be found in the $h^{x}_{AF}$ dependence. It indicates that within a range of $U$, the $z$-direction AF order is destroyed once the spin-orbit coupling is present. For $\lambda<0.25t$, when increasing $U$ further, we find the appearance of the $z$-term in the AF order eventually. However, for $\lambda\geq0.25t$, it has not been found up to $U=10t$. Thus, in the phase diagram we plot the white dashed-line separating the AF order with and without the $z$-term. The easy-plane AF order is the result of the interplay between the Hubbard interaction and the spin-orbit coupling. As is well known, the NN hopping will generate an isotropic AF Heisenberg term
$H_{1}=J_{1}\sum_{\langle ij\rangle}\mathbf{S}_{i}\cdot\mathbf{S}_{j}$ with $J_{1}=4t^{2}/U$ in the strong-coupling limit. Similarly, the NNN spin-orbit coupling generates
an anisotropic exchanging term
$H_{2}=J_{2}\sum_{\langle\langle ij\rangle\rangle}(-S^{x}_{i}S^{x}_{j}-S^{y}_{i}S^{y}_{j}+S^{z}_{i}S^{z}_{j})$~\cite{Rachel},
with $J_{2}=4\lambda^{2}/U$. Notice that the $z$ term in $H_{2}$ favors antiparallel alignment of the spins on the NNN sites, thus it will introduce a frustration to the NN AF correlation expressed by $H_{1}$.
On the other hand, the $xy$ term in $H_{2}$ favors a ferromagnetic alignment, so no frustration is introduced. As a result, the $H_{2}$ term coming from the spin-orbit coupling will suppress the $z$-term of the AF order.

\begin{figure}
  \centering
  \includegraphics[scale=1.0]{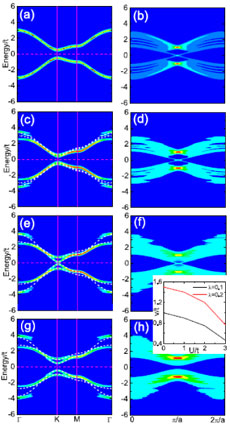}
  \caption{(color online) $A(\mathbf{k},\omega)$ for single-particle excitations in the bulk [Figs.(a), (c), (e) and (g)] and in a ribbon with the zigzag edges [Figs.(b), (d), (f) and (h)] as illustrated in Fig.\ref{fig1}(c) at $\lambda=0.1t$. The white dashed curves in Figs.(c), (e) and (g) are the mean-field fits discussed in the text. From the up to down figures, $U=0,2t,3t,4t$. The inset shows the $U$-dependence of the renormalized velocity of edge states at $\lambda=0.1t$ and $0.2t$.}
  \label{fig4}
\end{figure}

At another limit $U=0$, a TBI is expected to occur once the spin-orbit coupling is turned on~\cite{Kane}. The TBI is characterized by gapless edge states protected by the bulk gap opened by the spin-orbit
coupling in the single-particle spectrum. The spectral function of single particles is given by $A(\mathbf{k},\omega)=-2\mathrm{Im}G(\mathbf{k},\omega)/\pi$. The results for several $U$ at $\lambda=0.1t$ are presented
in Fig. \ref{fig4}, where the bulk bands are plotted along the lines shown in Fig. \ref{fig1}(b) and the edge states are calculated
from a ribbon with the zigzag edges [Fig. \ref{fig1}(c)]. For $U=0$, one can see that a bulk gap opens resulting from the spin-orbit coupling [Fig. \ref{fig4}(a)]. At the meantime, clear gapless edge states with sizeable spectral weights emerge [Fig. \ref{fig4}(b)]. These results reproduce perfectly the characters of a TBI~\cite{Kane}. In the presence of Hubbard interaction, the VCA captures exactly the short-range correlation effects by the exact diagonlization of the small clusters used to tile the lattice. We find that the bulk gap is reduced firstly and the edge states are stable against a weak $U$, as shown in Figs. \ref{fig4}(c) to (f). When $U$ is increased further, the bulk gap closes and the edge states disappear simultaneously. After that, a bulk gap with the character of the Mott gap occurs and no edge states reemerge anymore, as shown in Figs. \ref{fig4}(g) and (h). Thus, we determine the phase boundary where the TBI disappears by a criteria that the bulk gap closes and the edge states disappear. Combining with the results described above, we can conclude that the TBI phase will make transition to the SL phase when
$\lambda\leq0.125t$ and to the easy-$xy$ plane AF phase for $\lambda>0.125t$, as presented in the phase diagram of Fig. \ref{fig2}.

According to the bulk-boundary correspondence~\cite{Hasan}, the existence of gapless edge states depends on the topological class of the bulk band structure. The transition from TBI (topologically nontrivial state) to MI (toplogically trivial state) must undergo a gap closing process in the bulk. As far as we know, this process is demonstrated clearly for the first time by a systematic numerical calculation presented here.

A first attempt to understand the evolution of the spectrums in the KMH model is to include the AF order parameter $m^{A}=-m^{B}=|\langle n_{i\uparrow}-n_{i\downarrow}\rangle|$
($A$ and $B$ denote the sublattice in Fig. \ref{fig1}(a))~\cite{Rachel}, which is considered as a result of the electron correlations. This gives rise to the mean field dispersion given by $E(\mathbf{k})=\pm\sqrt{\varepsilon^{2}(\mathbf{k})+(\lambda-Um/2)^{2}}$, with $\varepsilon(\mathbf{k})$ the bare dispersion. When $Um/2=\lambda$, the spin-orbit gap closes. Then another gap $Um/2-\lambda$ with a character of the Mott gap opens up with the further increase of $U$. However, the evolution shown in Fig. \ref{fig4} exhibits a more complex behavior, namely both the valence and conduction bands around $K$ are split into two subbands. It implies that another interaction term is needed to be included. Comparing the numerical results for different $U$ and $\lambda$, we note that the band splitting around $K$ depends on $\lambda^{2}/U$. This is the exchange integral in $H_2$ coming from the second-order process of the spin-orbit interaction as described above. So, we rewrite $H_{2}$ as~\cite{Rachel} $H^{(2)}_{\lambda}=-(J_{2}/2)\sum_{\langle\langle ij\rangle\rangle}(a^{\dag}_{i\uparrow}a_{j\uparrow}-
a^{\dag}_{i\downarrow}a_{j\downarrow})(a^{\dag}_{j\uparrow}a_{i\uparrow}-a^{\dag}_{j\downarrow}a_{i\downarrow})$
and choose another parameter $\chi=\langle a^{\dag}_{i\uparrow}a_{j\uparrow}-a^{\dag}_{i\downarrow}a_{j\downarrow}\rangle$. By using $m^{A}$ and $\chi$ as adjustable parameters, we can give a fit to the numerical results, which is plot as white dashed lines in Fig. \ref{fig4}. This simple fit provides a possible understanding of the gap closing and reopening processes in the bulk.

Finally, let us discuss the possible effect of electron correlations on the edge states. As shown in the inset of Fig. \ref{fig4}, we notice a visible reduction of the velocity in helical Dirac fermions at the edge in the TBI phase. This renormalization arising from the two-particle scattering between the left and right moving modes due to electron correlations, which is allowed by the time reversal symmetry~\cite{Wu,Xu}.

In summary, we have investigated the interplay between the Hubbard interaction and the spin-orbit coupling in the Kane-Mele-Hubbard model with the variational cluster approach. We map a detail $U-\lambda$ phase diagram, in which the topological band insulator, the spin liquid, and the antiferromagnetic insulator are identified.  We have shown a nontrivial evolution of the bulk
bands in the topological quantum phase transition.

\begin{acknowledgments}
This work was supported by NSF-China and the MOST-China. XCX is also supported by US-DOE through the grant
DE-FG02-04ER46124.
\end{acknowledgments}

\end{document}